\documentclass[aps,pra,floatfix,twocolumn,floatfix,showpacs]{revtex4}

\usepackage{amsmath}

\begin{document}

\title{The sensitivity of hyperfine structure to nuclear radius and 
quark mass variation}

\author{T. H. Dinh$^1$, A. Dunning$^1$, V. A. Dzuba$^1$,
 and V. V. Flambaum$^{1,2}$}

\affiliation{$^1$School of Physics, University of New South Wales,
        Sydney, 2052, Australia\\
$^2$ New Zealand Institute for Advanced Study, Massey University,
 Auckland, New Zealand}
\date{\today}

\begin{abstract}
To search for the temporal variation of the fundamental constants
one needs to know dependence of atomic transition frequencies on these
constants. 
We study the dependence of the hyperfine structure of
atomic $s$-levels on nuclear radius and, via radius, on quark
masses. An analytical formula has been derived and tested by the numerical
relativistic Hartree-Fock calculations for Rb, Cd$^+$, Cs, Yb$^+$ and Hg$^+$.
The results of this work allow the use of the results of past and future
atomic clock experiments and quasar spectra measurements to put constrains
 on time variation of the quark masses.  

\end{abstract}


\maketitle

\section{Introduction}

Interest in the variation of the fundamental constants is motivated by
theories unifying gravity and other interactions.  
Indications that the fundamental constants might have varied slightly
from those of the distant past have been found in Big Bang
nucleosynthesis and quasar absorption spectra
 (see, e.g. reviews \cite{Uzan,Dzuba09,Flam07}).  
Most publications report only constraints on possible variations of
fundamental constants (see, e.g. reviews \cite{Flam07,Lea,karshenboim,
Karshenboim2000}).
Very stringent limits on the present time variation of the fundamental
constants have been obtained in the atomic clock experiments (see, e.g.
Refs. \cite{prestage,Marion2003,Bize2005,Peik2004,Bize2003,Fischer2004,
Peik2005,Peik2006,Fortier,Sr08,budker,budker1,Rosenband}). The majority of
recent work has been devoted to the variation of the fine structure
constant $\alpha$.  
However, the hypothetical unification of all interactions implies that
the variation of the dimensionless strong interaction parameter $X_q =
m_q/\Lambda_{QCD}$ (where $m_q=(m_u+m_d)/2$ is the average
current-quark mass and $\Lambda_{QCD}$ is the QCD scale) may be larger
than the variation of $\alpha$ (see for example \cite{Marciano,FlamWi} and the
references therein).   
In all intermediate  calculations it is convenient to assume that the strong
interaction scale $\Lambda_{QCD}$ does not vary, so we will speak of
the variation of masses (this means that we measure masses in units
of $\Lambda_{QCD}$).  
We will restore the explicit appearance of $\Lambda_{QCD}$ in the
final answers.

 
In a previous paper \cite{Thorium08} calculations  of the
sensitivity  of the nuclear radii to quark mass variation
were performed.  In the present paper we calculate the
dependence of hyperfine transition frequency on nuclear radius
(a preliminary approximate analytical result of this work was presented
in Ref. \cite{Thorium08}). Combining the results of the present
work and  Ref. \cite{Thorium08} we calculate the dependence
of the hyperfine structure on the quark masses and 
 the dimensionless strong interaction parameter $X_q =
m_q/\Lambda_{QCD}$.
 These calculations are needed to use the results of very
accurate atomic clock experiments to obtain constrains on the
variation of the fundamental constants.

The result of the present work is presented as a simple analytical formula.
To test this formula and improve the accuracy we have performed numerical
relativistic Hartree-Fock calculations for all atoms of experimental interest
Rb, Cd$^+$, Cs, Yb$^+$ and Hg$^+$ (and Tl in excited 7s state). They happened
to be atoms and ions with one s-wave electron above closed shells. One can use
 our analytical formula  for other atoms where the hyperfine structure is
 dominated by s-wave electrons.
 For other electrons the effect of the nuclear radius variation is small and
 may be neglected.

\section{Dependence of hyperfine transition frequency on nuclear
  radius and quark mass}

It has been found in Ref.~\cite{Thorium08} that the variation
of a nuclear radius $r_n$  can be related to variation of
quark mass by
\begin{equation}
  \frac{\delta r_n}{r_n} \approx 0.3  \frac{\delta
    m_q}{m_q},
\label{eq:rn}
\end{equation}
Numerical factor 0.3 in (\ref{eq:rn}) is approximately the same for all nuclei.
Then the variation of the frequency of a hyperfine transition $\omega_h$
may be presented as 
\begin{equation}
  \frac{\delta\omega_h}{\omega_h} \approx 0.3 k_{hr} \frac{\delta
    m_q}{m_q},
\label{eq:K}
\end{equation}
where 
\begin{equation}
  k_{hr} = \frac{\delta\omega_h/\omega_h}{\delta r_n/r_n}
\label{eq:Khr}
\end{equation}
 In this section we calculate
$k_{hr}$ using analytical and numerical approaches. 

To take into account finite nuclear size in the magnetic dipole
hyperfine structure (hfs)
Hamiltonian we approximate the nucleus by a uniformly magnetized
sphere. Then the Hamiltonian has a form
\begin{eqnarray}
  \hat H_{hfs} &=& - \frac{e}{c}\boldsymbol{\mu} \cdot \left[
    \mathbf{n} \times \boldsymbol{\alpha}\right] U(r) , \nonumber \\
    U(r)&=&\left\{ \begin{array}{ll}  	\frac{r}{r_n^3}, 	& ~r <
        r_{n}~  \\*  \frac{1}{r^2},  & ~r \geq r_{n}~ 
                   \end{array} \right.  .
\label{eq:hhfs}
\end{eqnarray}
 
Here $\mathbf{n}=\mathbf{r}/r$,$\boldsymbol{\alpha}$ is Dirac matrix,
$\boldsymbol{\mu}$ is 
nuclear magnetic moment and $r_n$ is nuclear radius.

For the analytical consideration we use a model of the nucleus where
the nuclear charge is considered to be uniformly distributed about a
sphere of radius $r_n$. Such a charge distribution corresponds to the
potential 
\begin{equation}
V(r)=\left\{ \begin{array}{ll}
    -\frac{Ze^2}{r_n}\left(\frac{3}{2}-\frac{r^2}{2 r_n^2}\right),
    & ~r < r_{n}~   \\*  -\frac{Ze^2}{r},  & ~r \geq r_{n}~  \end{array}
\right.
\label{eq:Vr}
\end{equation}
It is convenient to present the hyperfine frequency as 
 $\omega_h =\omega_0(1-\delta_h)$, where $\omega_0$ is the frequency
 at $r_n=0$ and $\delta_h$ describes the
change of the hfs frequency due to finite nuclear radius $r_n$. For the
 potential (\ref{eq:Vr}) the electron wave
 functions in the vicinity of the
nucleus can be found analytically and we obtain  the following approximate
expression for $\delta_h$ (see appendix for details):
\begin{equation}
\delta_h \approx \tfrac{72}{35}(Z r_n/a_B)^{2\gamma-1} ,
\label{eq:dha}
\end{equation}
where $\gamma=\sqrt{1-Z^2\alpha^2}$ and $a_B$ is the Bohr radius.
 Then we obtain
\begin{equation}
  k_{hr} =\frac{\delta\omega_h/\omega_h}{\delta r_n/r_n}= - \frac{(2\gamma-1)\delta_h}{1-\delta_h},
\label{eq:deltah}
\end{equation}

To check these results with a more accurate approach we calculate
atomic hfs constants using the relativistic Hartree-Fock method (see,
e.g.\cite{Dzuba84}). 
Relativistic Hartree-Fock Hamiltonian for atoms with one external
electron above closed shells can be written as
\begin{equation}
  \hat H^{HF}=c\mbox{\boldmath$\alpha$}\cdot{\bf p}+(\beta-1)m
  c^{2}+V_{nuc}(r)+V^{N-1} \ .
\label{eq:hf}
\end{equation}
Here $V_{nuc}(r)$ is nuclear potential and $V^{N-1}$ is the
self-consistent Hartree-Fock potential of the closed-shell atomic
core containing $N-1$ atomic electrons. Nuclear potential $V_{nuc}(r)$
is found by numerical integration of the Fermi-type distribution of
the nuclear charge. We assume the same electric and magnetic
radius. The same Hamiltonian (\ref{eq:hf}) is used for core and
valence states. The hfs frequencies are expressed via expectation
values of the hfs Hamiltonian (\ref{eq:hhfs}) over wave functions of
the valence electron calculated with the Hamiltonian
(\ref{eq:hf}). Note that neither core polarization nor correlation
effects are important for the {\em relative} change of the hfs
frequency for s-wave energy levels since the polarization and correlation
corrections are dominated by the matrix elements of the hyperfine interaction
between the s-wave orbitals (the hyperfine
 matrix elements between the p,d,... orbitals are significantly smaller).
We have tested this conclusion using the full-scale many-body calculations
for the Cs hyperfine structure using approach developed in our work
 \cite{Dzuba89}. With the accuracy $\sim$ 1\% the polarization
and correlation corrections do not change the relative value of the effect
of the variation (to avoid misunderstanding we should note that the
 polarization and correlation corrections change the hyperfine structure
 constant by $\sim$ 40\%). 

 To find the change of the hyperfine frequency due to the change
of nuclear radius we perform calculations for at least
two different nuclear radii and than calculate derivative numerically
\begin{equation}
  \frac{\delta\omega_h}{\delta r_n} = \frac{\omega_h(r_n+\delta
    r_n)-\omega_h(r_n-\delta r_n)}{2\delta r_n}.
\label{eq:der}
\end{equation}
The values of $k_{hr-cal}$ found from the Hartree-Fock calculations are
presented in Table~\ref{tab:Khr}. Using (\ref{eq:deltah}) we can
express $\delta_h$ in a form similar to (\ref{eq:dha})
\begin{equation}
\delta_h \approx C(Z r_n/a_B)^{2\gamma-1} ,
\label{eq:dha1}
\end{equation}
where $C$ is a fitting factor found from a comparison of the results
of calculations with the formula (\ref{eq:dha1}). The values of $C$
for some atoms are presented in Table~\ref{tab:Khr}. Its variation
from atom to atom is small and average value is 1.995. Using $r_n =
1.1A^{\frac{1}{3}}$ fm and the total number of nucleons
$A \approx 2.5 Z$ leads to the formula
\begin{equation}\label{hyperfine3}
 \delta_h = 1.995 \times (2.8 \cdot 10^{-5} Z^{4/3})^{2 \gamma-1} 
\end{equation} 
which can be used for any medium or heavy atom. Note that this result agrees
with the purely analytical result eq. (\ref{eq:dha})  to the 
accuracy of few per cent.

\begin{table*}
\caption{The sensitivity of the hyperfine transition frequency to
  variation of the nuclear radius,
analytical ($k_{hr}$) from eq.~(\ref{eq:deltah},\ref{hyperfine3}) and
numerical ($k_{hr-cal}$) results.}  
\label{tab:Khr}
\begin{ruledtabular}
\begin{tabular}{l c c c c c c c}
 Atom or Ion  & $^{87}_{37}$Rb & $^{111}_{48}$Cd$^+$ & $^{133}_{55}$Cs & $^{171}_{70}$Yb$^+$ 
& $^{199}_{80}$Hg$^+$ & $^{205}_{81}$Tl(7s) & $^{233}_{87}$Fr\\\hline
$ k_{hr}$         &  -0.010   & -0.017   & -0.024  & -0.048  & -0.077  & -0.081  &  -0.111  \\
$ k_{hr-cal}$         &  -0.0096   & -0.0171   & -0.0242  & -0.0492 & -0.0778  & -0.0798  &  -0.1082  \\
$ C$              &   1.9514   &  2.0034   &  2.0338  &  2.0500 &  2.0050  &  1.9623  &   1.9563  \\
\end{tabular}
\end{ruledtabular}
\end{table*}


Now we can use the expression (\ref{eq:K}) to calculate the sensitivity of
 $\omega_h$ to the quark mass due to the variation
of nuclear radius (parameter
$k_{hq}=0.3 \cdot k_{hr}$). All results are displayed  in Table \ref{tab:vary}.
Also presented in Table \ref{tab:vary} are the parameters $K_{rel}$,
$k_{\mu}$ and $k= k_{\mu}+k_{hq}$. $K_{rel}$ is the sensitivity of the
hyperfine structure to variation of $\alpha$  derived from the results of the
atomic many-body calculations \cite{Dzuba99}. Parameter
$k_{\mu}$ is the sensitivity of the nuclear magnetic moment to quark
mass calculated in Ref. \cite{Flambaum06}.

It is convenient to present the final results using the ratio of the hyperfine
energy  $E_h=\hbar \omega_h$ to the atomic unit of
 energy $E_a = m_ee^4/\hbar^2$. Atomic experiments always measure the ratio
of two atomic frequencies. The atomic unit of energy $E_a$ cancels
out in such ratios. Following Ref. \cite{Flambaum06} we 
 define the parameter $V$ through the relation
\begin{equation} \label{eqn:V def}
\frac{\delta V}{V} = \frac{\delta (E_h/E_a)}{E_h/E_a}.
\end{equation}
Then one can use  Table \ref{tab:vary} to find the dependence of the hyperfine
transition frequencies on the variation of the fundamental constants using
the following formula from Ref. \cite{Flambaum06} :
\begin{equation} \label{eqn:V} 
V =  \alpha^{2+K_{rel}} \left(\frac{m_q}{\Lambda_{QCD}}\right) ^{k}
  \frac{m_e}{m_p} 
\end{equation}
A number of  limits on variation of $V$ from different experiments
are presented in Ref. \cite{Flambaum06}. These results give the best present
time limits on the variation of $m_q/\Lambda_{QCD}$.
For example, for $\omega_h(^{87} \mbox{Rb})/\omega_h(^{133} \mbox{Cs})$, we have 
\begin{equation}
\label{RbCs}
X(\mbox{Rb/Cs})=\frac{V(^{87} \mbox{Rb})}{V(^{133} \mbox{Cs})} =
 \alpha^{-0.49} \left(\frac{m_q}{\Lambda_{QCD}}\right)
 ^{-0.021}
\end{equation}
and the result of measurements by \cite{Bize2005} can be presented as a limit on the variation of X:
\begin{equation}
\label{limit}
\frac{1}{X(\mbox{Rb/Cs})} \frac{dX(\mbox{Rb/Cs})}{dt}
 = (-0.5 \pm 5.3)\times 10^{-16}/\mbox{yr}. 
\end{equation}
Using a very stringent limit on the variation of $\alpha$
obtained using  our calculations \cite{dzuba} and measurements in 
Ref. \cite{Rosenband},
\begin{equation}
\label{limit1}
\frac{1}{\alpha} \frac{d\alpha}{dt}
 = (-1.6 \pm 2.3)\times10^{-17}/\mbox{yr}, 
\end{equation}
we may find the variation
of  $X_q=m_q/\Lambda_{QCD}$ from  eqs. (\ref{RbCs}) and (\ref{limit}):
\begin{equation}
\label{limit2}
\frac{1}{X_q} \frac{d X_q}{dt}
 = (0.3 \pm 2.5)\times 10^{-14}/\mbox{yr}.
\end{equation} 
Note that the effect of the variation may be enhanced by 2-3 orders of
 magnitude in a number of molecules where the hyperfine splitting is
 approximately equal to an interval between the rotational levels
 \cite{Flambaum}.
\begin{table*}
\caption{The sensitivity of the hyperfine structure to variation of $\alpha$ (parameter $K_{rel}$) 
and to the quark mass/strong interaction scale $m_{q}/\Lambda_{QCD}$ (parameter $k=k_{\mu}+k_{hq}$)}.
\label{tab:vary}
\begin{ruledtabular}
\begin{tabular}{l c c c c c }
 Atom or Ion  & $^{87}_{37}$Rb & $^{111}_{48}$Cd$^+$ & $^{133}_{55}$Cs & $^{171}_{70}$Yb$^+$ 
& $^{199}_{80}$Hg$^+$ \\\hline
$ K_{rel}$  &  0.34  & 0.6    & 0.83   & 1.5    & 2.28   \\
$ k_{\mu}$  & -0.016 &  0.125 &  0.009 & -0.085 & -0.088 \\
$ k_{hq}$   & -0.003 & -0.005 & -0.007 & -0.014 & -0.023 \\
$ k $       & -0.019 &  0.120 &  0.002 & -0.099 & -0.111 \\
\end{tabular}
\end{ruledtabular}
\end{table*}


\acknowledgments

This work was supported in part by the Australian Research Council
and Marsden grant.

\section*{Appendix}

Analytical approach presented here is very similar to those used
in Ref.~\cite{Ginges}. To simplify all expressions we use atomic
units $\hbar=e=m_e=1, c=1/\alpha$ in the Appendix.

We use an electron wave function in the form
\begin{equation}
\psi_\kappa^{m_j}(\mathbf {r})= \frac{1}{r}\left(\begin{array}{c}
    f_\kappa(r)\chi_\kappa^{m_j}(\mathbf{\hat r})\\ 
    i g_\kappa(r)\chi_{-\kappa}^{m_j}(\mathbf{\hat r})
                   \end{array} \right) 
\end{equation}
where $\chi_{\kappa}^{m_j}$ are spherical spinors. For an s-orbital
the initial terms of a power series solution of the radial wave
functions inside the nucleus are 
\begin{align}  
f_n = & a\, x \left[1-\left(\frac{3}{8}\frac{Z^2}{c^2}+\frac{1}{2}Z
    r_n\right)x^2+\dots\right]\label{gnuc} \\ 
g_n = & -\frac{a\,Z}{2
  c}x^2\left[1-\left(\frac{1}{5}+\frac{9}{40}\frac{Z^2}{c^2}+\frac{3}{10}Z
    r_n\right)x^2+\dots\right]\label{fnuc} 
\end{align}
where $x=r/r_n$. Only those terms explicit in (\ref{gnuc}) and
(\ref{fnuc}) will be retained for further calculation. The external
radial wavefunctions take the form of Bessel functions 
\begin{align}  
f_e =&
\left(\gamma+\kappa\right)\left(J_{2\gamma}\left(y\right)+b\,Y_{2\gamma}\left(y\right)\right)\nonumber\\   
&-\frac{y}{2}\left(J_{2\gamma-1}\left(y\right)+b\,Y_{2\gamma-1}\left(y\right)\right)\\
g_e =& \frac{Z}{c}\left(J_{2\gamma}\left(y\right)+b\,Y_{2\gamma}\left(y\right)\right)\label{fex}
\end{align}
where, $\kappa =(-1)^{j-l+1/2}(j+1/2)$,
\begin{equation}
\gamma=\sqrt{\kappa^2-\alpha^2Z^2},    \quad   y=\sqrt{8Z\,r}
\end{equation}
As the binding energy of the electron is small in comparison to the
potential energy it has been neglected in the calculation of
(\ref{gnuc}-\ref{fex}). The constants $a$ and $b$ are found such that
the wave functions remain continuous at $r=r_n$. For small $r_n$ they
may be approximated by 
\begin{align}
a=\frac{(2 Z r_n)^\gamma}{\Gamma(2
  \gamma)}&\left(-\frac{1}{5}(2\gamma+3)+\frac{3}{80}\frac{Z^2}{c^2}(3\gamma+7)\right.\nonumber\\ 
&\left.+\frac{1}{20}Z r_n(3\gamma+7)\right)^{-1}\\
b=-\frac{a \pi (2 Z r_n )^\gamma}{\Gamma(1+2\gamma)}&
\left(-\frac{1}{5}(2\gamma-3)+\frac{3}{80}\frac{Z^2}{c^2}(3\gamma-7)\right.\nonumber\\ 
&\left.+\frac{1}{20}Z r_n(3\gamma-7) \right)
\end{align}
As we shall only consider relative changes in the hyperfine
interaction normalization of the wave functions is not necessary. The
first order correction to the energy is simply the s-wave expectation
value of the hyperfine interaction which may be shown to be 
\begin{equation}
\omega_h=\left<s|\hat H_{hfs}|s\right>=k \int U(r)\,f\,g\,dr
\end{equation}
where k is a constant \cite{Schwartz55}. In the limit of zero nuclear radius this becomes
\begin{equation}
\omega_0=k\int_0^\infty \frac{1}{r^2}f_0 g_0 dr=\frac{k\, Z^2(1-2\kappa)}{c (\gamma-4 \gamma^3)}
\end{equation}
where $f_0\!=\!f_e , g_0\!=\!g_e (r_n\!=\!0)$. We define a relative
change in the hyperfine interaction $\delta_h$ by
$\omega_h=\omega_0(1-\delta_h)$. 

\begin{align}
\delta_h=&-\frac{k}{\omega_0}\int_0^{r_n}\left(\frac{r}{r_n^3}f_n
  g_n-\frac{1}{r^2}f_0 g_0\right)dr\nonumber\\ 
&-\frac{k}{\omega_0}\int_{r_n}^\infty \frac{1}{r^2}\left(f_0 \delta
  g_e+g_0 \delta f_e\right)dr 
\end{align}
where $\delta g =g_e-g_0$ and $\delta f =f_e-f_0$. For small $r_n$ these integrals result in
\begin{align}
\delta_h=-\frac{a^2\gamma(4\gamma^2-1)}{Z r_n}\left(
  \frac{1}{35}-\frac{13}{1008}\frac{Z^2}{c^2}-\frac{13}{756}Z
  r_n\right.\nonumber\\ 
\left.+\frac{1}{640}\frac{Z^4}{c^4}+\frac{1}{240}\frac{Z^3 r_n}{c^2}+\frac{1}{360}Z^2 r_n^2\right)\nonumber\\
+\frac{(2Z r_n )^{2\gamma-1}2\gamma(1+\gamma)(1+2\gamma)}{3\,
  \Gamma(1+2\gamma)^2}+\frac{b(4\gamma^2-1)}{3\pi Z r_n} 
\end{align}
Keeping only leading in $Zr_n$ terms reduces this to
\begin{equation}
\label{final}
\delta_h \approx \tfrac{72}{35}(Z r_n)^{2\gamma-1}
\end{equation}
Differentiation of $\omega_h$ with respect to $r_n$ and rearranging yields
\begin{equation}
\frac{\delta \omega_h}{\omega_h}=-\frac{(2\gamma-1)\delta_h}{1-\delta_h}\frac{\delta r_n}{r_n}
\end{equation}

\end{document}